# Dynamic Pattern of Finite-Pulsed Beams inside One-dimensional Photonic Band Gap Materials

Li-Gang Wang and Shi-Yao Zhu

Department of Physics, Zhejiang University, Hangzhou, 310027, China
and Faculty of Science and Technology, University of Macau, Macau
and Department of Physics, Hong Kong Baptist University, Hong Kong

## Abstract

The dynamics of two-dimensional electromagnetic (EM) pulses through one-dimensional photonic crystals (1DPC) has been theoretically studied. Employing the time expectation integral over the Poynting vector as the arrival time [Phys. Rev. Lett. 84, 2370, (2000)], we show that the superluminal tunneling process of EM pulses is the propagation of the net forward-going Poynting vector through the 1DPC, and the Hartman effect is due to the saturation effect of the arrival time (smaller and smaller time accumulated) of the net forward energy flow caused by the interference effect of the forward and the backward field (from the interfaces of each layer) happened in the region before the 1DPC and in the front part of the 1DPC.

Superluminal tunneling of an electromagnetic (EM) pulse propagating through a barrier has been a subject of considerably theoretical attention [1-4] because of the "mysterious" tunneling time delay independent of the barrier length, known as the Hartman effect [5]. This phenomenon was first noticed by MacColl more than 70 years ago [6]. It was confirmed by experimental results [2-3] that tunneling time delay of a light wave packet (or a photon) though a one-dimensional photonic crystal (1DPC) is in good agreement with the prediction of the group delay by means of the stationary phase method. Other experiments by Enders and Nimtz [4] also showed that EM pulses travel with superluminal group velocity as they tunnel through an "undersized" waveguide. These observations lead to various arguments proposed to explain the mechanism of superluminal tunneling and why superluminal group velocity does not violate causality. The argument based on the "pulse reshaping" was widely believed that the tunneling pulse comes from the small early front of the incident pulse [1]. Japha and Kurizki [7] further argued that "predominantly destructive interference between accessible causal paths" is responsible for the tunneling attenuation, superluminal group delay and pulse reshaping. Most recently Winful [8, 9] presented a different description by introducing the concepts of energy storage and release to resolve the mystery, and he thought that "distortionless tunneling of electromagnetic pulses through a barrier is a quasistatic process" and "the envelopes of the transmitted and reflected fields can adiabatically follows the incident pulse with only a small delay that originates from energy storage". However, the question of how the tunneling pulse is carved from the incident pulse is still unsatisfying [10] (or not understood [11]), or say, remains hot controversial [12-13].

In this paper, we present the dynamics of the EM pulses through a finite 1DPC by using the finite-pulsed beams. To our best knowledge, it is the first time to describe the superluminal tunneling from the view point of energy transport by using the Poynting vector. As mentioned in Ref. [14], for the EM energy flow, the field is solely responsible for the energy transport [15]; this is true even if the total energy of a pulse is composed of the energy in the field and the energy stored in the medium. Meanwhile, only the Poynting vector can be measured directly. Therefore, we use a time expectation integral [14] over the net forward-going Poynting vector as the arrival time. It is found that the EM tunneling process is the propagation of the net forward-going Poynting vector (i. e. the net energy flow), and the tunneling time delay is directly originated from the arrival time of the net forward energy flow. This may help us understand the dynamics of the superluminal tunneling process and the "Hartman effect".

Consider the dynamic evolution of a two-dimensional (2D) pulsed beam with TE polarization passing through a finite 1DPC used in several tunneling experiments [2-3]. As in Fig. 1(a), for TE polarization, the electric field is in the $x$ direction, $\vec{E} \equiv E_x(y,z;t)\vec{x}$, and the magnetic field is in the $y-z$ plane, $\vec{H} \equiv H_y(y,z;t)\vec{y} + H_z(y,z;t)\vec{z}$, where $\vec{x}$, $\vec{y}$ and $\vec{z}$ are the unit vectors in the $x$, $y$ and $z$ directions, respectively. The incident electric field of the pulsed beam is given by the following 2D Fourier integral,

$$E_x^{(i)}(y,z,t) = \frac{1}{2\pi} \iint \tilde{E}(k_y, z_0, \omega) \exp\left[ik_z(z-z_0) + ik_y y\right] \exp\left[-i\omega t\right] dk_y d\omega, \ (z<0), \quad (1)$$

where $k_z = \sqrt{k^2 - k_y^2}$ for $k^2 > k_y^2$, otherwise $k_z = i\sqrt{k_y^2 - k^2}$ in the vacuum, and the function $\tilde{E}(k_y, z_0, \omega) = A \exp[-\tau^2(\omega - \omega_0)^2 / 4] \exp[-W_y^2(k_y - k_{y0})^2 / 4]$ denotes the initial spectrum with a half-width of $1/\tau$ and angular spectrum of the incident 2D pulsed beam with Gauss-shaped profile at the initial plane $z = z_0 < 0$, with $k_{y0} = k\sin\theta$, $W_y = W\sec\theta$, $W$ the half-width of the pulsed beam at waist, $\theta$ the incident angle, $k = \omega/c$, $\omega_0$ the center angular frequency and $c$ the speed of light in vacuum. The reflected and transmitted electric fields from the 1DPC are, respectively, expressed as

$$E_x^{(r)}(y,z,t) = \frac{1}{2\pi} \iint r(k_y, \omega) \tilde{E}(k_y, z_0, \omega) \exp\left[-ik_z(z+z_0) + ik_y y\right] \exp\left[-i\omega t\right] dk_y d\omega$$

$$(z < 0), \quad (2)$$

$$E_x^{(t)}(y,z,t) = \frac{1}{2\pi} \iint t(k_y, \omega) \tilde{E}(k_y, z_0, \omega) \exp\left[ik_z(z-z_0-L) + ik_y y\right] \exp\left[-i\omega t\right] dk_y d\omega$$

$$(z > L). \quad (3)$$

We have assumed that both sides of the 1DPC are the vacuum. Here $r(k_y, \omega)$ and $t(k_y, \omega)$ are, respectively, the reflection and transmission coefficients of the 1DPC directly calculated from the transfer matrix method [16], and $L = \sum_{i=1}^{N} d_i$ is the total length of the 1DPC (where $d_i$ is thickness of the $i$th layer and $N$ is the total layer number). The spectrum of the magnetic field is related to that of the electric field via the relation $\vec{H}(y,z,\omega)=\frac{1}{i\mu\omega}\nabla\times\vec{E}(y,z,\omega)$. Therefore, the temporal-spatial evolution of the magnetic field can be obtained by $\vec{H}(y,z,t)=\frac{1}{\sqrt{2\pi}}\int\vec{H}(y,z,\omega)d\omega$. For any transversal wave vector $k_y$ at $\omega$, the total electric field and the $y$ component of the total magnetic field at two positions $z$ and $z+\Delta z$ inside the $j$ th layer can be related by a transfer matrix [16]

$$M_j(k_y,\Delta z,\omega)=\begin{pmatrix}\cos(k_z^j\Delta z) & i\sin(k_z^j\Delta z)/q_j \\ iq_j\sin(k_z^j\Delta z) & \cos(k_z^j\Delta z)\end{pmatrix}. \qquad (4)$$

Here $k_z^j=\sqrt{\varepsilon_j\mu_j k^2-k_y^2}$ is the $z$ component of the wave vector in the $j$ th layer for $\varepsilon_j\mu_j k^2>k_y^2$, otherwise $k_z^j=i\sqrt{k_y^2-\varepsilon_j\mu_j k^2}$; $q_j=k_z^j/(\mu_j k)$. Thus the total electric field and the $y$ and $z$ components of the total magnetic field at position $z$ inside the $j$ th layer of the 1DPC can be expressed by [16]

$$E_x(k_y,z,\omega)=A^{(i)}(k_y,0,\omega)\{[1+r(k_y,\omega)]Q_{11}(k_y,z,\omega)+q_0[1-r(k_y,\omega)]Q_{12}(k_y,z,\omega)\}, \quad (5)$$

$$cH_y(k_y,z,\omega)=A^{(i)}(k_y,0,\omega)\{[1+r(k_y,\omega)]Q_{21}(k_y,z,\omega)+q_0[1-r(k_y,\omega)]Q_{22}(k_y,z,\omega)\}, \quad (6)$$

$$cH_z(k_y,z,\omega)=p_jA^{(i)}(k_y,0,\omega)\{[1+r(k_y,\omega)]Q_{11}(k_y,z,\omega)+q_0[1-r(k_y,\omega)]Q_{12}(k_y,z,\omega)\}. \quad (7)$$

Here $p_j=-\frac{k_y}{\mu_j k}$, and $A^{(i)}(k_y,0,\omega)=\tilde{E}(k_y,z_0,\omega)\exp[-ik_z z_0]$ are the function of the spectrum and angular spectrum of the 2D pulsed beam arriving at the plane $z=0$, and $Q_{\alpha\beta}(k_y,z,\omega)$ ( $\alpha,\beta=1,2$ ) are the elements of the matrix $Q(k_y,z,\omega)=M_j(k_y,\Delta z,\omega)\prod_{i=1}^{i=j-1}M_i(k_y,d_i,\omega)$, where $z=\Delta z+\sum_{i=1}^{i=j-1}d_i$. Finally we can obtain the dynamic evolutions of the total EM field inside the finite 1DPC as follows,

$$E_x(y,z,t)=\frac{1}{2\pi}\int\int E_x(k_y,z,\omega)\exp(ik_y y)\exp[-i\omega t]dk_y d\omega, \qquad (8a)$$

$$cH_i(y,z,t) = \frac{1}{2\pi} \iint cH_i(k_y,z,\omega)\exp(ik_y y)\exp\left[-i\omega t\right]dk_y d\omega , \qquad (8b)$$

where the subscripts $i = y,z$ denote the $y$ and $z$ components of the magnetic field.

Recently Winful [8, 9] used the concept of energy storage to investigate the tunneling process using the total energy density. As mentioned by Diener [17], the total energy density is composed of two parts: one is the *propagating* energy and another is the *non-propagating* energy stored in the medium; the energy stored in the medium cannot propagate directly and is not responsible for the energy transport. Another kind of energy carried by the standing wave also *cannot* propagate [18]. In order to unfold the dynamics of the pulsed beam inside the finite 1DPC, the Poynting vector $\vec{S}(y,z,t)$ (the net energy flow) is used here. The change of $\vec{S}(y,z,t)$ in the space-time domain directly indicates the dynamics of energy transport during superluminal tunneling process. In our case, $\vec{S}(y,z,t)$ is given by

$$\vec{S}(y,z,t) = \frac{c}{4\pi}\text{Re}[\vec{E}(y,z,t) \times \vec{H}^*(y,z,t)] = s_z \vec{z} + s_y \vec{y} , \qquad (9)$$

where $s_z = \frac{1}{4\pi}\text{Re}[E_x(y,z,t)cH_y^*(y,z,t)]$ , $s_y = -\frac{1}{4\pi}\text{Re}[E_x(y,z,t)cH_z^*(y,z,t)]$ , and its magnitude is given by $\left|\vec{S}(y,z,t)\right| = \sqrt{s_z^2 + s_y^2}$ and its direction is given by the angle $\vartheta = \arctan(s_y/s_z)$ , $\vartheta \in [0,2\pi)$ . In the region $z < L$ , $\vec{S}(y,z,t)$ is the result of interference between the forward and the backward fields. Here we define $\vec{S}(y,z,t)$ with positive $s_z$ to be the net forward-going energy flow (NFEF), and define $\vec{S}(y,z,t)$ with negative $s_z$ to be the net backward-going energy flow (NBEF). Then *the arrival time* of the net energy flow along the $+z$ direction [at point $(y,z)$] defined by the time expectation integral (or "gravity center") of the NFEF (which is directly responsible for the tunneling pulse as discussed below) is given by [14]

$$<t>_{(y,z)} = \hat{u} \cdot \int_{-\infty}^{\infty} t\vec{S}(y,z,t)dt \Big/ \hat{u} \cdot \int_{-\infty}^{\infty} \vec{S}(y,z,t)dt , \qquad (10)$$

where the unit vector $\hat{u}$ refers to the direction in which the net energy flow is detected (normal to a detector surface). In our case we are interested in the arrival time of the NFEF, so $\hat{u}$ is along the +z direction and the integral is only taken over the value of $\vec{S}(y,z,t)$ with positive $s_z$. Here we would like to emphasize that due to the reflection, the NFEF (or NBEF) is the result of the interference between the forward and backward fields.

Consider a symmetric 1DPC: $(AB)^m A$, where $m$ is an integer. All layers are nonmagnetic ($\mu \equiv 1$) and are characterized by their dielectric constants $\varepsilon_{A,B}$, and their thicknesses satisfy $\sqrt{\varepsilon_A} d_A = \sqrt{\varepsilon_B} d_B = \lambda_{pc}/4$ (where $\lambda_{pc}$ is the mid-gap wavelength of the 1DPC at normal incidence). The parameters are $\varepsilon_A = 4.84$, $\varepsilon_B = 1.96$, and $\lambda_{pc} = 3 \, \text{mm}$ (corresponding to the angular frequency $\omega_{pc}/2\pi = 100 \, \text{GHz}$). The incident 2D pulsed beam is normal incident from $z_0 = -100 \, \text{mm}$ on the finite 1DPC with $\theta = 0$, $W = 30 \, \text{mm}$, $\tau = 0.1 \, \text{ns}$ and $\omega_0 = \omega_{pc}$. Figure 1(b) shows a broad photonic band-gap structure with $m = 9$ and the spectral position of the incident 2D pulsed beam. It's clear seen that the spectrum and angular spectrum of the incident pulsed beam is narrow and completely confined within the photonic band-gap. So it is expected that the tunneling pulse is with superluminal group delay. In our case, the group delay (defined by $\tau_g \equiv d\varphi/d\omega$ [2], where $\varphi$ is the phase of the transmission coefficient) is about 6.67 **ps** which is much shorter than the time delay $\tau_v = L/c = 27.44$ **ps** passing through the same distance in the vacuum.

Figure 2(a)-2(h) show the typically dynamic patterns of the net energy flow for the 2D pulsed beam tunneling through the finite 1DPC at different times: at the initial time, the net energy flow is forward-going through the 1DPC because the forward field is predominated. As the pulse approaches to the 1DPC [near the time of the pulse peak arriving at the incident end $z = 0$, i. e., near $t = (1/3) \, \text{ns}$], the net energy flow, starting from the off-axis region ($|y| > 0$), begins to change from the NFEF to the NBEF, see Fig. 2(a)-(e). Note that at the position where the net energy flow changes from the NFEF to the NBEF, the net energy flow may align along the $+y$ (or $-y$) direction for $y > 0$ (or $y < 0$), i.e. $s_z = 0$. At these points the forward $z$-component energy flow associated with the forward field cancels the backward $z$-component energy flow associated with the backward electric field. Note that the energy carried by the standing wave or stored in the medium, which corresponds to the non-propagating energy, is not responsible for the tunneling pulse. Once the net energy flow becomes the NBEF in a region, the NBEF has no more net contribution for the output pulse, because the backward field is predominated thereafter [18]. It is interesting that there appears a clear division (i.e., $s_z = 0$) between the NFEF and the NBEF inside the 1DPC and this division will be pushed forward and will be close to the exit end as the tunneling pulse leaves away from the 1DPC. Figures 2 (e) and (f) show that on left side of the division there is only the NBEF which leads to the reflected pulse and on right side of the division only the NFEF propagates forward and through the exit end to form the tunneling pulse. Therefore, **the tunneling pulse comes directly from the propagation of the NFEF**.

Now we are going to explain why the tunneling time is independent of the layers of the 1DPCs, i.e. the optical Hartman effect. We use Eq. (10) to calculate the arrival time of the NFEF and to trace the tunneling process. For the points on the $z$ axis, $\vec{S}(y \equiv 0, z, t)$ only has the $z$ component. As discussed above, only the NFEF is responsible for the tunneling pulse, while the NBEF has no net contribution to the transmitted pulse [18]. Figure 3(a) shows the arrival times at the points on the z axis for the pulsed beam tunneling through the 1DPCs with different total layers. It is seen that the arrival times at the exit ends of the different 1DPCs are nearly the same [see the open triangles on each curves or on inset (i) of Fig.3 (a)], i.e., the Hartman effect. From Fig. 3(a), we see that, before the incident end, the time needed for propagating a unit distance is shorter than that for the vacuum, and becomes smaller and smaller as the pulse approaches the incident end, i.e., a saturation for the arrival time as the pulse approaching the incident surface. The change from NFEF to NBEF happens nearly at the same time at different $z$ (<0) points, see curves ①-④ in Fig. 3(b). As $m$ increases (i.e., deep photonic band gap and large reflectivity), the transition time of the net energy flow from the NFEF to the NBEF occurs earlier [see open cycles on inset (i) of Fig. 3(a)]. The arrival time at $z=0$ decreases (although very small) with increasing $m$ [see the solid cycles on insert (i)]. The saturation is almost independent of $m$ for large $m$ before the incident surface ($z<0$). In the front layers of the 1DPC, the saturation continues with almost zero time for propagating through them (for example from ④-⑤). As $m$ increases, the saturation effect immerges deeper and deeper into the 1DPCs. How deep the saturation will be depends on $m$ (the reflectivity), see the curves for $m=9,12,15$ in Fig. 3(a). The higher reflectivity, the deeper into the 1DPC the saturation will be. The saturation can be understood in the following. As the pulse approaching the 1DPC, more and more energy is stored [9] into polarization of the medium and the forms of the standing wave [18] (the non-propagating energy) due to reflection at surfaces between layers, which leads to that the "gravity center" of the effective NFEF moves quickly forward in space before and in the front of the 1DPC (i.e., shorter and shorter time spent). The effectively forward-propagating energy (the NFEF) decreases greatly until the stored energy will gradually release into the NBEF (forming the reflected pulse) and small amount into NFEF (forming the output pulse) [see Fig. 3(d)]. As the pulse enters the rear part of the 1DPC, the time needed to propagate through a unit distance increases and finally becomes longer comparing with the vacuum case, because the reflected field becomes weaker and weaker due to less and less layers. The times in passing through the last a few layers are almost independent of $m$ [see inset (ii) of Fig. 3(a)]. Because the transition time from the NFEF to the NBEF happens later and later, see curves ⑤-⑨ in Fig 3(c), the arrival time (the "gravity center" of the NFEF) increases gradually to be identical with the tunneling time. That is to say, the "gravity center" of the NFEF moves very slowly in the rear part of the 1DPC. In fact, since the backward field becomes weaker and weaker in the rear part, the energy stored in the standing wave and the medium will become smaller and smaller, which results in that the NFEF gradually tends to be identical with the energy flow of the tunneling pulse as it reaches the exit end (see the positive parts in the curves ⑤-⑨).

Figure 4 shows the spatial-temporal evolutions for the forward and backward, and the total electric fields at points along the z axis of the 1DPC with m=9. Obviously, the forward electric field does decay inside the 1DPC, like the evanescent wave. Therefore, the forward energy inside the 1DPC becomes less and less due to the reflection. The interference between the forward and backward fields lead to the strong standing wave before and in the front part of the 1DPC [see Fig. 4 (c)]. In the space-time domain, there is a clear division between the NFEF and the NBEF (see the blue dashed curve in Fig. 4 (c)), which leads to the saturation effect before and in the front part of the 1DPC [see Fig. 3 (a)] as discussed in above. In the space-time domain above the division, the backward field is dominated so that the net energy flow is the NBEF, which has no net contribution to the tunneling pulse. Only is the NFEF directly responsible for the tunneling pulse, see the arrows in Fig. 4 (c). *Therefore, the tunneling time is the result of the arrival time of the NFEF.*

In conclusion, we have shown the dynamic patterns of EM pulses through the finite 1DPC. From the point view of net forward energy flow, see Eq. (10), we found that the superluminal tunneling process of EM pulses is the propagation of the net forward-going Poynting vector through the 1DPC, and the Hartman effect is due the saturation effect of the arrival time of the NFEF happened *in the region before the 1DPC and in the front part of the 1DPC* due to the strong reflection from all the surfaces between layers. The contribution to the arrival time for propagating through the region before the 1DPC and in the front part of it is very small, and the main contribution comes from propagating through the rear part of the 1DPC.

This research was supported by the NSFC under Contact No. 10547138, and NSFC/05-06/01.

# FIGURE CAPTIONS

FIG. 1. (a) Schematic of the 1DPC; (b) Photonic band-gap structure of $(AB)^9 A$, and the normalized spectrum and angular spectrum of the incident 2D pulsed beam (top right inset).

FIG. 2. The dynamic patterns of the net energy flow for the 2D pulsed beam tunneling through the 1DPC of $(AB)^9 A$ at times (a) $t = 0.333333$ ns, (b) $t = 0.336$ ns, (c) $t = 0.3367$ ns, (d) $t = 0.337$ ns, (e) $t = 0.34$ ns, and (f) $t = 0.5$ ns.

FIG. 3. (Color Online). (a) The arrival time of the NFEF at the different points on the z axis for the pulse tunneling through the 1DPCs with different $m$. Inclined dashed line denotes the arrival time of the NFEF for the vacuum case. Vertical thin dashed line denotes the incident end and the open triangles denote the arrival times of the NFEF at the exit ends of different 1DPCs. Inset (i) shows the dependence of the arrival times of the NFEF at point (0,0) (solid cycles) and the exit ends (open triangles) of different 1DPCs on the period number $m$, and open cycles denote the transition times of the net energy flow from the NFEF to the NBEF at point (0,0); inset (ii) shows the (nearly overlapped) arrival times in the rear parts of the different 1DPCs (from the above to below curves for $m = 15,12,9,6$), and note that the z axis is rescaled by the coordinates of the exit points. In (b) and (c), curves ①-⑨ show the temporal behaviors of $s_z$ at points ①: $z = -40$ mm, ②: $z = -15$ mm, ③: $z = -5$ mm, ④: $z = 0$ (incident end), ⑤: $z = 4.12$ mm, ⑥: $z = 5.10$ mm, ⑦: $z = 6.01$ mm, ⑧: $z = 7.08$ mm and ⑨: $z = 8.23$ mm (the exit) on the z axis for the 1DPC with $m = 9$ [corresponding to the points ①-⑨ in (a)]. In (d) it shows time dependence of different kinds of energies (distributed along the z axes) on the time in the tunneling process of the 1DPC with $m = 9$, curve A, the NFEF, curve B, the NBEF, curve C, the storage energy, and curve D, the energy flow in the case of the pulse propagating in the vacuum.

FIG. 4. (Color Online). Spatial-temporal evolutions for the intensities of (a) the forward electric field, (b) the backward electric field, and (c) the total electric field at points along the z axis before and inside the 1DPC with m=9. In (c) the arrows denote the directions of the energy flow; blue dashed curve is the division of the NFEF and the NBEF.

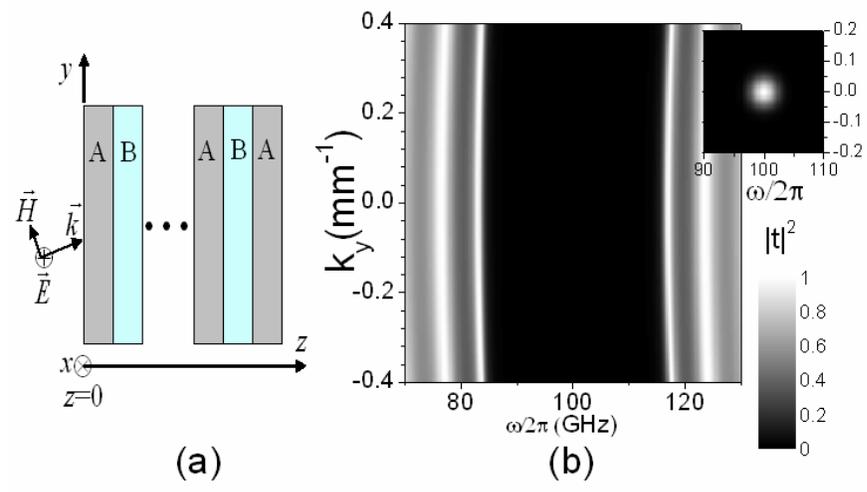

(a)

(b)

FIG.1

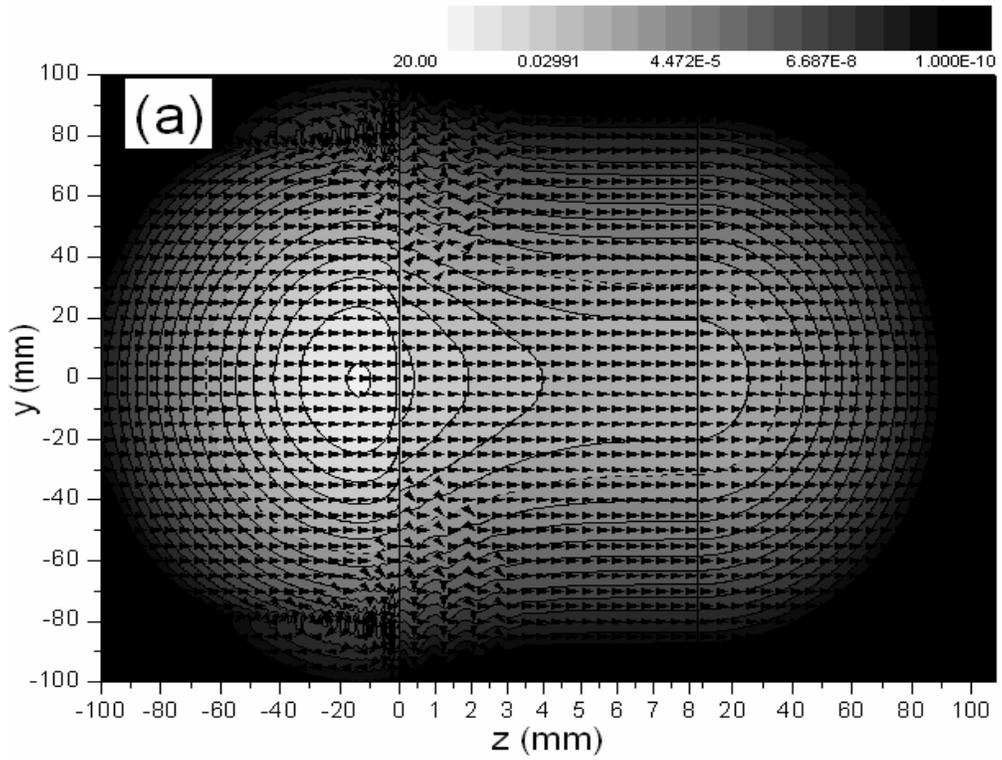

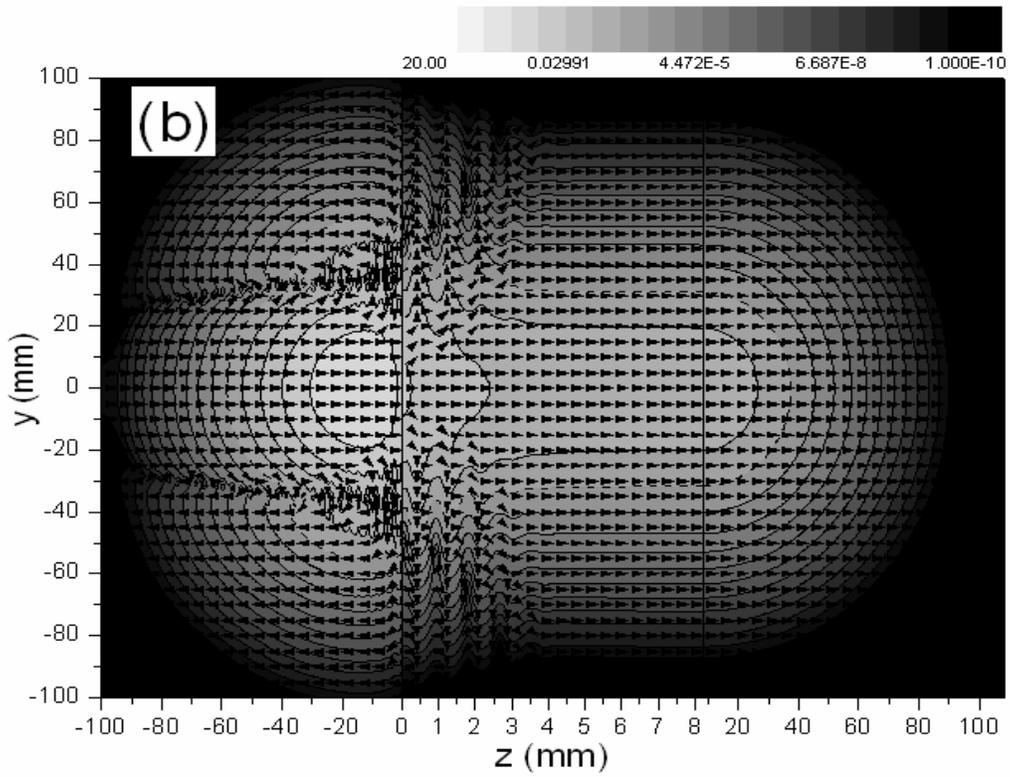

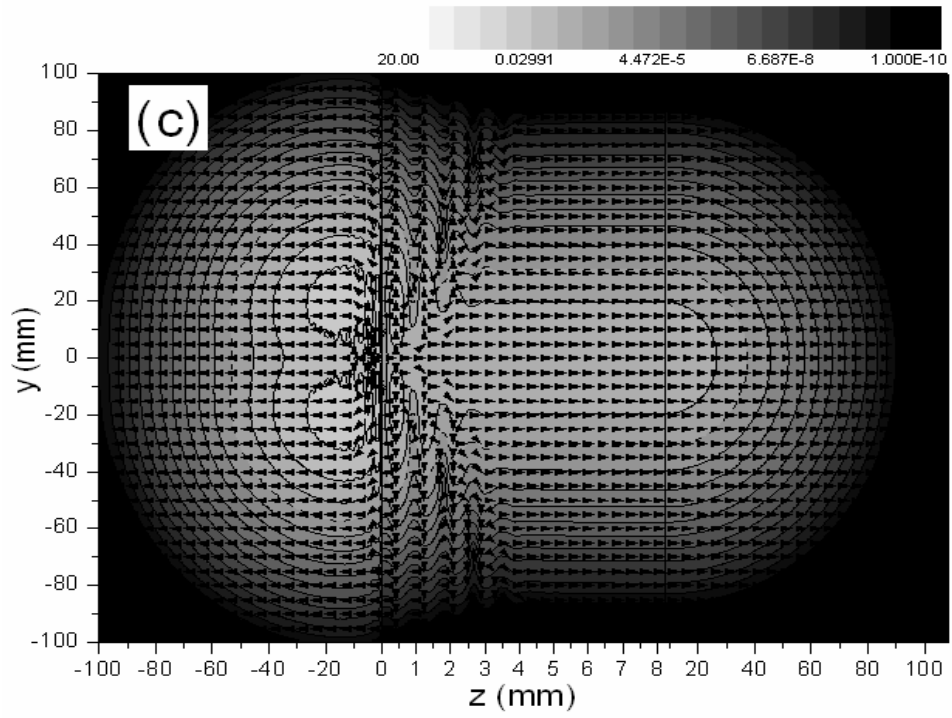

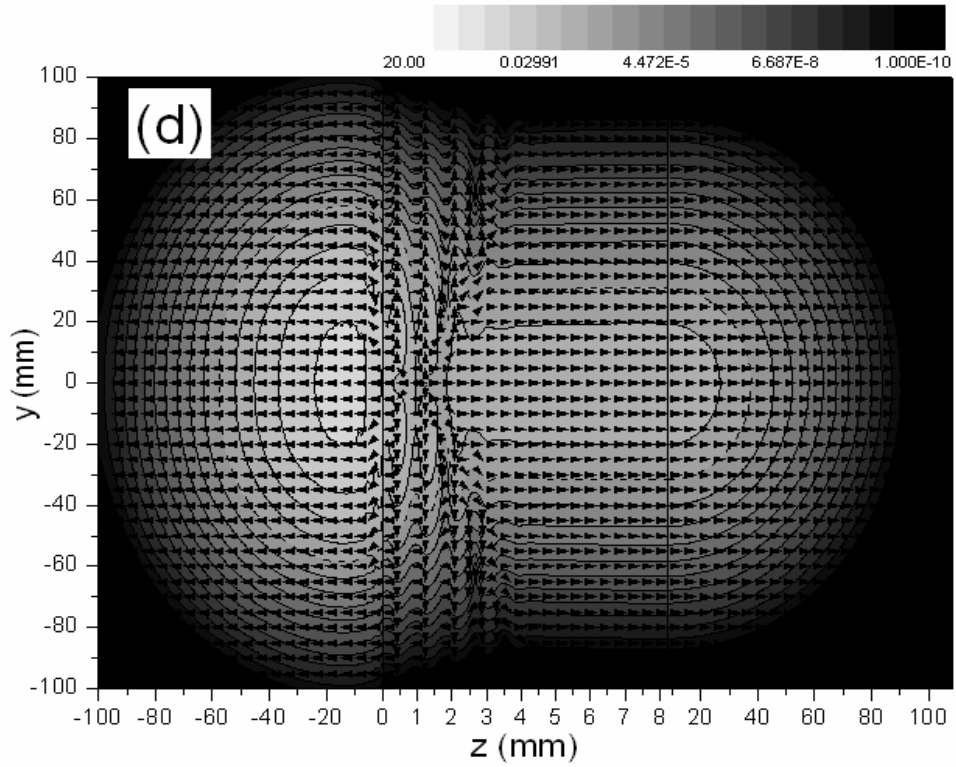

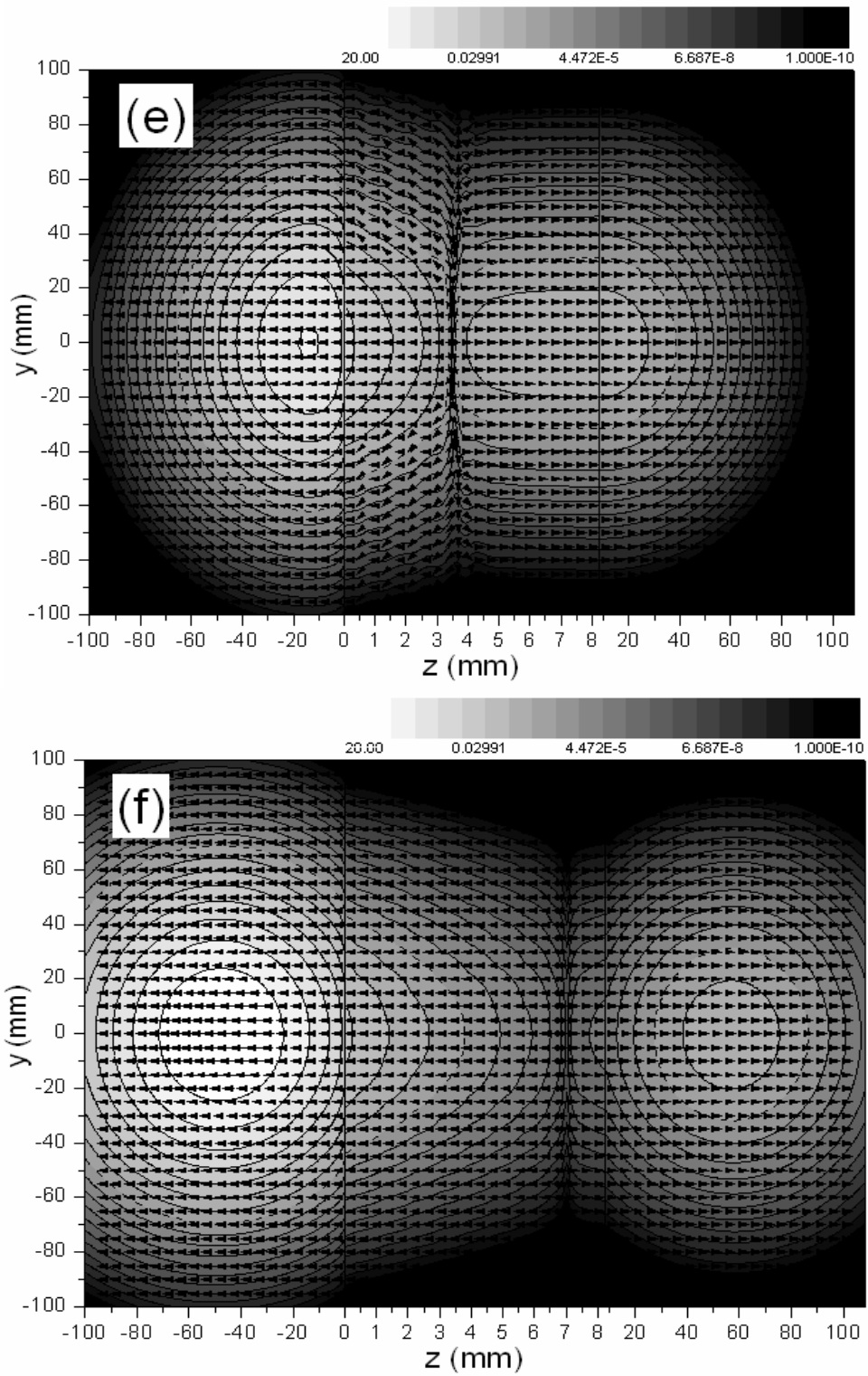

FIG. 2.

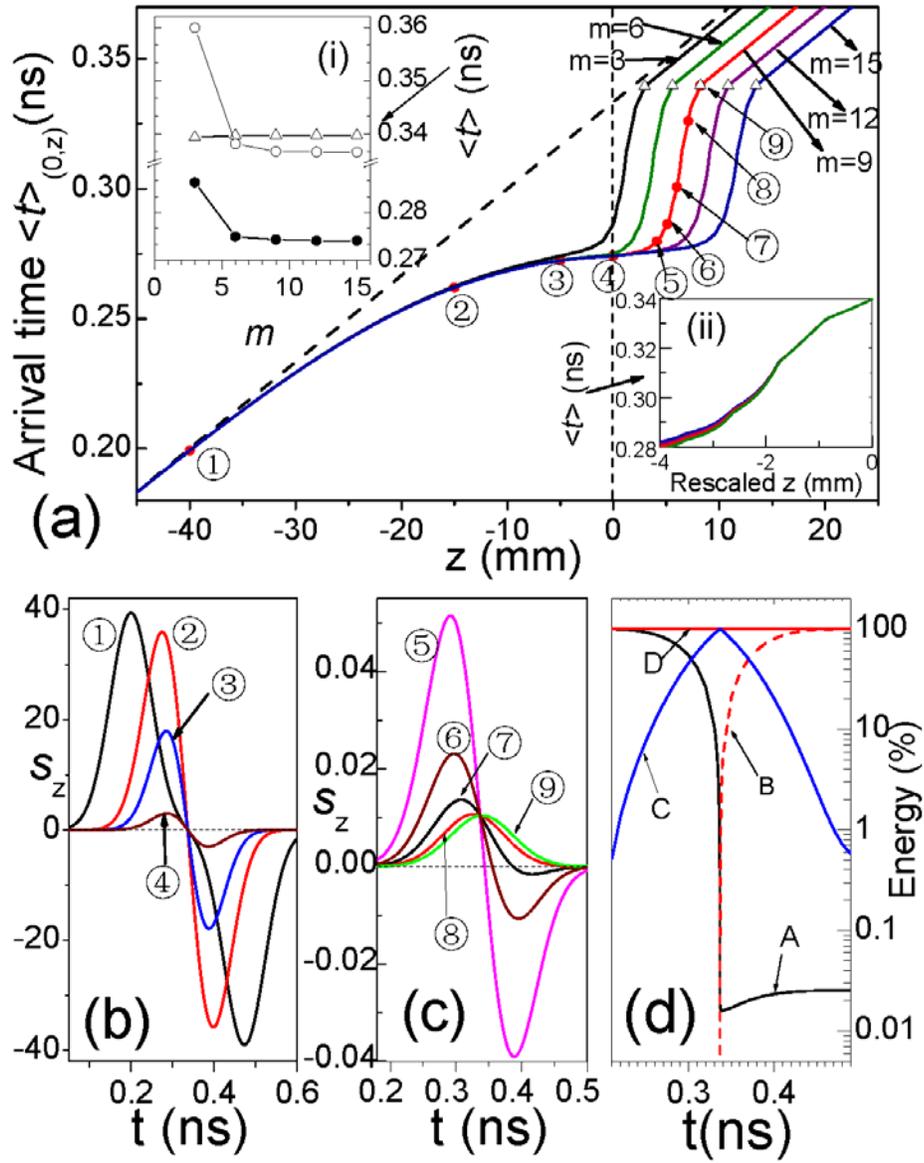

FIG. 3.

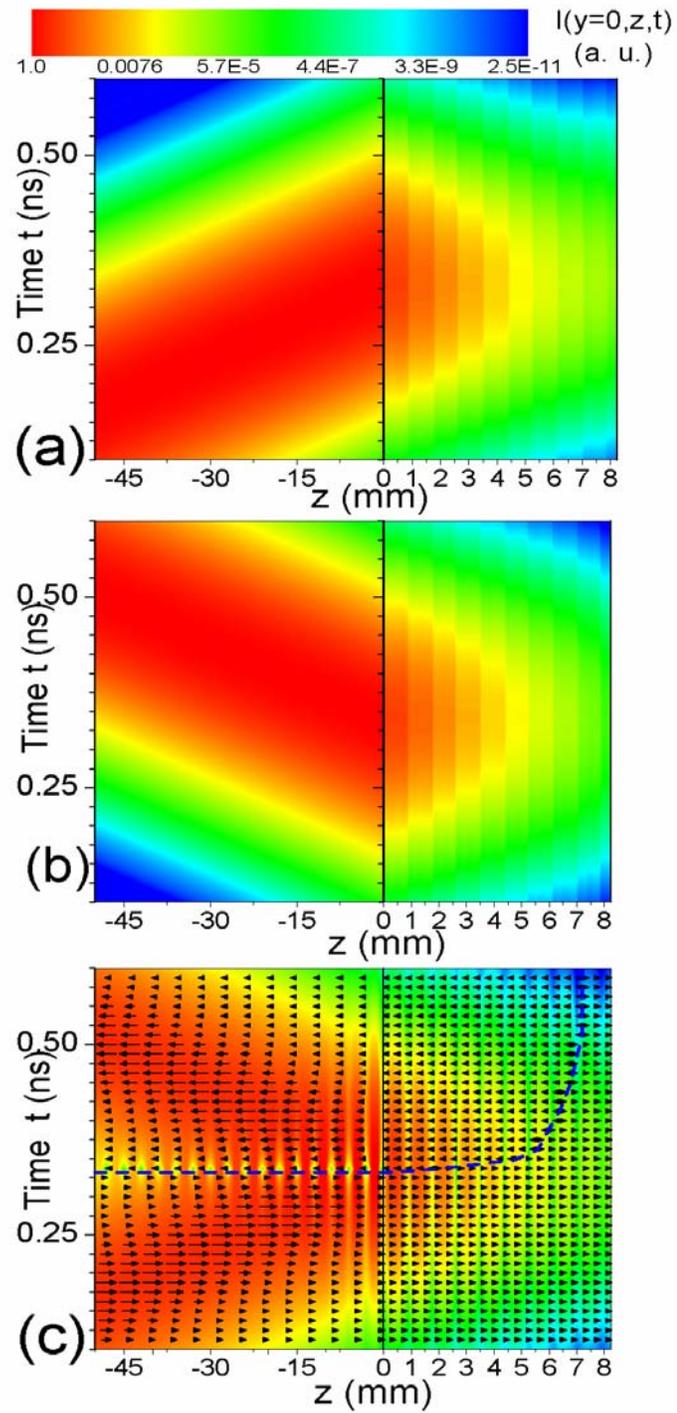

FIG. 4